\documentclass[twoside,twocolumn]{article}
\usepackage{xr}

\usepackage[sc]{mathpazo} 
\usepackage[T1]{fontenc} 
\usepackage{lmodern} 
\usepackage{microtype} 

\usepackage[english]{babel} 

\usepackage{graphicx}
\usepackage{color}
\usepackage[top=2cm,bottom=2cm,left=1.5cm,right=1.5cm,columnsep=20pt]{geometry} 
    {\large\begin{scshape} Computational Details\end{scshape}%
    \par\medskip\normalsize}%
    {}%
    {\large\begin{scshape} Acknowledgement\end{scshape}%
    \par\medskip\normalsize}%
    {}%
    {\large\begin{scshape} Supporting Information\end{scshape}%
    \par\medskip\normalsize}%
    {}%
\usepackage{multirow} 

\usepackage{amsmath} 

\usepackage[normal, small,labelfont=bf,up,textfont=it,up]{caption} 
\usepackage{booktabs} 

\usepackage{lettrine} 

\usepackage{enumitem} 
\setlist[itemize]{noitemsep} 

\usepackage{abstract} 

\usepackage{titlesec} 
\renewcommand\thesection{\Roman{section}} 
\renewcommand\thesubsection{\roman{subsection}} 
\titleformat{\section}[block]{\large\scshape\centering}{\thesection.}{1em}{} 
\titleformat{\subsection}[block]{\large}{\thesubsection.}{1em}{} 

\usepackage{fancyhdr} 
\usepackage{titling} 

\usepackage{hyperref} 

\usepackage{kotex}

\def\bea{\begin{eqnarray}}
\def\eea{\end{eqnarray}}
\def\ben{\begin{equation}}
\def\een{\end{equation}}
\def\benu{\begin{enumerate}}
\def\enu{\end{enumerate}}

\def\bei{\begin{itemize}}
\def\eei{\end{itemize}}
\def\beit{\begin{itemize}}
\def\eit{\end{itemize}}
\def\benu{\begin{enumerate}}
\def\enu{\end{enumerate}}

\def\sss{\scriptscriptstyle\rm}

\def\x{_{\sss X}}
\def\c{_{\sss C}}

\def\xc{_{\sss XC}}

\def\sph_int{ {\int d^3 r}}

\usepackage[table,xcdraw]{xcolor}
\usepackage{amsmath}
\usepackage{graphicx}
\usepackage{chemformula}
\usepackage{float}
\usepackage{dblfloatfix}
\usepackage{comment}
\usepackage{cancel}

\newcolumntype{C}{>{\centering\arraybackslash}p{4.5em}}

\def\d{_{\sss D}}
\def\f{_{\sss F}}

\def\dSt{$\Delta\langle \hat{S}^2 \rangle$}
\def\dStp{$\Delta\langle \hat{S}^2 \rangle \%$}
\def\St{$\langle \hat{S}^2 \rangle$}
\def\S{$\rm \tilde{S}$}
\def\rtscan{r$^2$SCAN}
\def\dde{$\Delta E_{\sss D}$}
\def\fe{$\Delta E_{\sss F}$}

\DeclareMathOperator{\WTMAD}{WTMAD-2}

\DeclareMathOperator{\mae}{MAE}
\DeclareMathOperator{\mar}{MAR}

\title{DC-DFT for Open Shells: How to Deal with Spin Contamination}

\author{%
\textsc{Hayoung Yu$^a$, Suhwan Song$^a$, Seungsoo Nam$^a$, Kieron Burke$^b$, and Eunji Sim$^a$\thanks{esim@yonsei.ac.kr}}\\ 
\normalsize $^a$Department of Chemistry, Yonsei University, 50 Yonsei-ro Seodaemun-gu, Seoul 03722, Korea \\
\normalsize $^b$Department of Chemistry, University of California, Irvine, CA 92697, USA \\ 
}
\date{}

\usepackage{subcaption}

\usepackage{xr}
\makeatletter

\newcommand*{\addFileDependency}[1]{
\typeout{(#1)}
\@addtofilelist{#1}
\IfFileExists{#1}{}{\typeout{No file #1.}}
}\makeatother

\renewcommand{\maketitlehookd}{%
\begin{abstract}
\noindent

Density functional theory (DFT) is widely used to predict chemical properties,
but its accuracy is limited by functional approximations and their
approximate self-consistent densities. Density-corrected DFT (DC-DFT) is the study of the errors
due to densities and Hartree-Fock DFT (HF-DFT) uses HF densities 
to improve energetics. With increasing use of HF-DFT, 
the question of how to address strong spin
contamination in the HF calculation becomes increasingly important. 
We compare two different open-shell HF densities across 13
different DFT functionals and two DC-DFT methods.
For significant spin contamination, ROHF densities outperform
UHF densities by as much as a factor of 3, depending on
the energy functional, and ROHF-DFT improves over self-consistent DFT for most of the tested functionals.
We refine the DC(HF)-DFT algorithm, recommending ROHF-DFT in cases of severe spin contamination.

\end{abstract}
}

\begin{document}

\maketitle


\sf


Density functional theory (DFT) is a method for calculating the properties of electronic systems using
the electron density as the basic variable. 
Given the exact exchange-correlation (XC) functional, the exact density is found in the Kohn-Sham (KS) equations, 
and exact energies and associated properties can be extracted.
In practical calculations, the exact energy functional is unknown, 
and DFT is performed with density functional approximations (DFA) using approximated functionals
and their self-consistent (sc) densities. 

\begin{figure}[!htbp]
\centering
\includegraphics[width=\columnwidth]{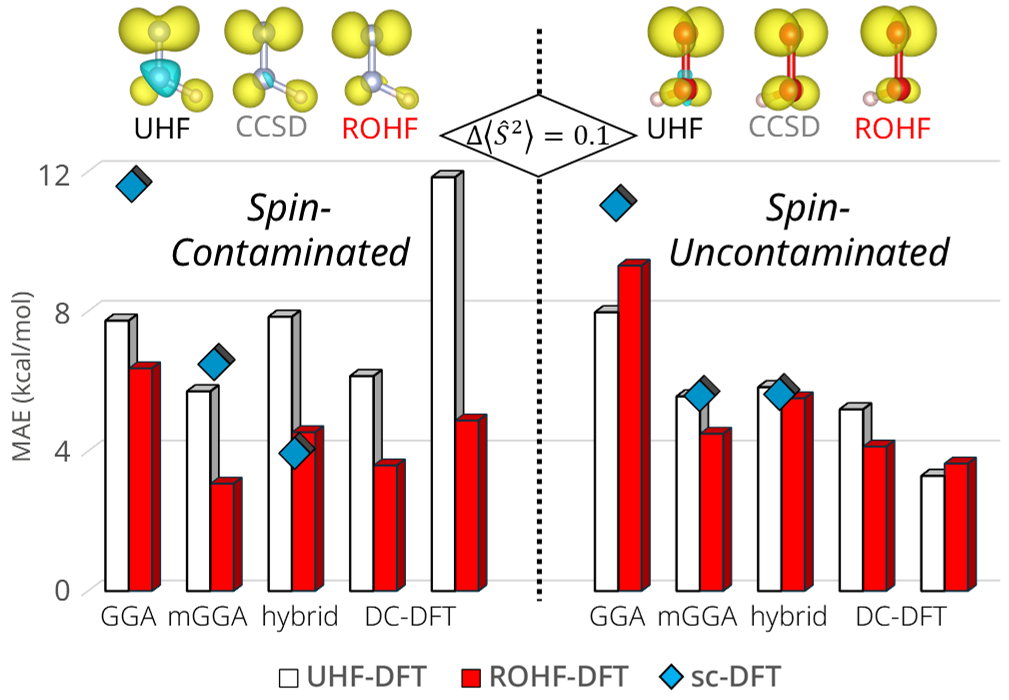} 
\caption{
Mean absolute errors (MAE, kcal/mol) of open-shell subsets of GMTKN55 database using various self-consistent (sc-)/UHF-/ROHF-DFT methods. 
Spin-contaminated cases are on the left and spin-uncontaminated cases are on the right. 
The plotted functionals are PBE (GGA), \rtscan{} (mGGA), PBE0 (hybrid), HF-\rtscan{}-DC4, and BL1p (DC-DFT).
Detailed numbers are given in Table~\ref{tab:15fnls_table} below. 
Top: Spin density plot examples for UHF, ROHF, and CCSD of molecules HNN (left) and HOO (right).
}
\label{fig:firstpic}
\end{figure}
Density-corrected DFT (DC-DFT)\cite{KSB13} provides a theoretical framework
with which to analyze the origin of errors in any DFT calculation.\cite{SVSB22, SSVB22}
In many cases, 
such as stretched NaCl and HO$\cdot$Cl$^{-}$ radicals, the
Hartree-Fock (HF) 
density is sufficiently close to a high-level density, yielding energies that
are significantly better than with sc-densities.\cite{NSSB20}
This method, to use HF density in place of sc-densities, is called HF-DFT.
While HF-DFT is not always a synonym for DC-DFT, 
HF-DFT itself has shown remarkable performance 
emerging as an extremely useful practical procedure.\cite{TCCLweb}
The cases where HF-DFT showed remarkable success include
pure water and aqueous systems,\cite{DLPP21,DSBP22,SVKY22}
electron and hole polaron defects,\cite{RCH22}
crystal polymer conformational energies,\cite{RBH22}
making and breaking of internal hydrogen bonds,\cite{SM21}
torsional barriers,\cite{NCSB21} 
electron affinity,\cite{KSB11} 
dissociation energy curves of heteronuclear molecules,\cite{NSSB20,KPSS15} 
radical ions in aqueous solution,\cite{KSB14} 
spin gaps of Fe(II) complexes,\cite{SKSB18}
halogen and chalcogen binding energies,\cite{KSSB19}
reaction barrier heights,\cite{SSVB22,JS08,VPB12}
etc. 
HF-DFT not only works for the energetics, but also provides sufficient accuracy similar to sc-DFT in various molecular properties.\cite{MA23}

In the past,
when using the HF density for the purpose of DC-DFT, i.e., DC(HF)-DFT,
the use of HF densities in HF-DFT has been limited to cases without severe spin contamination
(typically \St{} error above 0.1), based on the argument that spin-contaminated HF densities are unlikely to be
accurate.\cite{SVSB22} 
(In those cases, DC(HF)-DFT simply reverts to sc-DFT).   
Figure~\ref{fig:firstpic} shows that for many functionals (not all), there
is little difference between restricted open-shell HF (ROHF) and unrestricted HF (UHF) if spin contamination is small or zero (right-hand side), but
 dramatic improvements occur in spin-contaminated (SC) cases (left-hand side). 
The greatest improvement on spin-contaminated cases is for functionals
designed with DC-DFT principles, i.e., whose energetics are trained with density-driven
errors removed.

We stress this does {\em not} mean (a) ROHF gives better energetics than UHF, (b) that 
breaking of symmetries in sc-DFT calculations is good or bad, and (c) that
even the ROHF spin-densities are somehow 'better' than those of UHF.   All it means is
that they yield better energetics when several approximate fuctionals are evaluated
on those spin densities.

The rest of this paper essentially gives context to Fig.~\ref{fig:firstpic}. 
First, the background of DC-DFT and spin contamination in unrestricted (U-) calculations are briefly explained, 
and results and discussion for different densities applied in various functionals follow.  
Finally, the practical application of DC-DFT is revised.


\noindent
{\em Background:} DC-DFT emphasizes several important points to be considered when designing approximate
functionals and analyzing their performance.\cite{KSB13,KSB14}
The total error $\Delta E$ in any sc-DFT calculation can be written as
\begin{equation}
    \Delta E = \tilde{E}[\tilde{n}] - E[n] = \Delta E\f + \Delta E\d
\end{equation}
where $E[]$ is the exact total energy functional and $n$ is the exact electron density for the given system,
while tilde indicates their approximate counterparts. 
This total error can be split into two parts, 
the functional error (\fe) and the density-driven error (\dde);
\begin{equation} 
\begin{split}
\label{eq:DC-DFT}
\Delta E\f &= \tilde{E}[n] - E[n], \\
\Delta E\d &= \Delta E - \Delta E\f = \tilde{E}[\tilde{n}] - \tilde{E}[n].
\end{split}
\end{equation}
In most DFT calculations, the error is dominated by the functional contribution.
But, in many well-characterized situations, the density-driven error can be
unusually large (called 'abnormal' calculations), and use of the exact
density significantly reduces the total error.
Empirical approximations, fitted only to the total error, risk unintended
cancellation between these error sources, leading to lower accuracy for cases
where the cancellation does not occur.

\begin{table}[!htp]
\centering
\tiny
\begin{tabular}{rr|rrr|rrr|}
\rowcolor[HTML]{D9D9D9} 
\multicolumn{2}{l|}{\cellcolor[HTML]{D9D9D9}}                       & \multicolumn{3}{c|}{\cellcolor[HTML]{D9D9D9}SC (46)}                                                                                                    & \multicolumn{3}{c|}{\cellcolor[HTML]{D9D9D9}SU (384)}                                                                                                   \\
\rowcolor[HTML]{D9D9D9} 
\multicolumn{2}{l|}{\cellcolor[HTML]{D9D9D9}}                       & \multicolumn{1}{c}{\cellcolor[HTML]{D9D9D9}sc-} & \multicolumn{1}{c}{\cellcolor[HTML]{D9D9D9}UHF-} & \multicolumn{1}{c|}{\cellcolor[HTML]{D9D9D9}ROHF-} & \multicolumn{1}{c}{\cellcolor[HTML]{D9D9D9}sc-} & \multicolumn{1}{c}{\cellcolor[HTML]{D9D9D9}UHF-} & \multicolumn{1}{c|}{\cellcolor[HTML]{D9D9D9}ROHF-} \\ \hline
\rowcolor[HTML]{FFFFFF} 
\multicolumn{1}{l}{\cellcolor[HTML]{FFFFFF}GGA}         & BLYP      & 7.60                                            & 19.36                                            & 8.69                                               & 7.72                                            & 5.98                                             & 6.85                                               \\
\rowcolor[HTML]{FFFFFF} 
\multicolumn{2}{r|}{\cellcolor[HTML]{FFFFFF}PBE}                    & 16.65                                           & 16.65                                            & 7.23                                               & 9.89                                            & 5.20                                             & 6.13                                               \\
\rowcolor[HTML]{FFFFFF} 
\multicolumn{2}{r|}{\cellcolor[HTML]{FFFFFF}PW91}                   & 16.74                                           & 15.96                                            & 7.13                                               & 10.17                                           & 5.24                                             & 6.13                                               \\
\rowcolor[HTML]{FFFFFF} 
\multicolumn{2}{r|}{\cellcolor[HTML]{FFFFFF}RPBE}                   & 13.83                                           & 18.44                                            & 6.70                                               & 8.23                                            & 5.28                                             & 5.98                                               \\
\rowcolor[HTML]{FFFFFF} 
\multicolumn{2}{r|}{\cellcolor[HTML]{FFFFFF}revPBE-D3(BJ)$^\dag$}   & 14.44                                           &                                                  &                                                    & 8.14                                            &                                                  &                                                    \\ \hline
\rowcolor[HTML]{FFFFFF} 
\multicolumn{1}{l}{\cellcolor[HTML]{FFFFFF}mGGA}        & M06-L     & 13.72                                           & 14.47                                            & 8.09                                               & 6.63                                            & 5.12                                             & 5.54                                               \\
\rowcolor[HTML]{FFFFFF} 
\multicolumn{2}{r|}{\cellcolor[HTML]{FFFFFF}TPSS}                   & 14.67                                           & 15.04                                            & 6.54                                               & 8.22                                            & 5.54                                             & 5.71                                               \\
\rowcolor[HTML]{FFFFFF} 
\multicolumn{2}{r|}{\cellcolor[HTML]{FFFFFF}SCAN}                   & 14.00                                           & 7.32                                             & 4.78                                               & 6.97                                            & 4.54                                             & 4.09                                               \\
\rowcolor[HTML]{FFFFFF} 
\multicolumn{2}{r|}{\cellcolor[HTML]{FFFFFF}r$^2$SCAN}              & 12.40                                           & 8.84                                             & 4.49                                               & 6.70                                            & 4.42                                             & 4.23                                               \\
\rowcolor[HTML]{FFFFFF} 
\multicolumn{2}{r|}{\cellcolor[HTML]{FFFFFF}SCAN-D3(BJ)$^\dag$}     & 14.18                                           &                                                  &                                                    & 7.01                                            &                                                  &                                                    \\ \hline
\rowcolor[HTML]{FFFFFF} 
\multicolumn{1}{l}{\cellcolor[HTML]{FFFFFF}hybrid}      & B3LYP     & 9.96                                            & 16.66                                            & 5.77                                               & 6.53                                            & 4.96                                             & 5.09                                               \\
\rowcolor[HTML]{FFFFFF} 
\multicolumn{2}{r|}{\cellcolor[HTML]{FFFFFF}TPSSh}                  & 11.48                                           & 13.75                                            & 5.08                                               & 6.97                                            & 5.34                                             & 5.25                                               \\
\rowcolor[HTML]{FFFFFF} 
\multicolumn{2}{r|}{\cellcolor[HTML]{FFFFFF}M06}                    & 7.50                                            & 17.18                                            & 6.92                                               & 4.24                                            & 3.97                                             & 4.25                                               \\
\rowcolor[HTML]{FFFFFF} 
\multicolumn{2}{r|}{\cellcolor[HTML]{FFFFFF}PBE0}                   & 8.48                                            & 13.07                                            & 5.24                                               & 5.52                                            & 3.43                                             & 3.71                                               \\
\rowcolor[HTML]{FFFFFF} 
\multicolumn{2}{r|}{\cellcolor[HTML]{FFFFFF}M06-2X}                 & 2.67                                            & 23.44                                            & 7.12                                               & 3.34                                            & 3.14                                             & 3.53                                               \\
\rowcolor[HTML]{FFFFFF} 
\multicolumn{2}{r|}{\cellcolor[HTML]{FFFFFF}$\omega$B97X-V$^\dag$}   & 5.82                                            &                                                  &                                                    & 1.51                                            &                                                  &                                                    \\
\rowcolor[HTML]{FFFFFF} 
\multicolumn{2}{r|}{\cellcolor[HTML]{FFFFFF}DSD-BLYP-D3(BJ)$^\dag$} & 4.97                                            &                                                  &                                                    & 3.51                                            &                                                  &                                                    \\ \hline
\rowcolor[HTML]{FFFFFF} 
\multicolumn{2}{r|}{\cellcolor[HTML]{FFFFFF}HF-r$^2$SCAN-DC4}       & 12.60$^\ddag$                                  & 8.97                                             & 4.69                                               & 6.71$^\ddag$                                    & 4.31                                             & 4.13                                               \\
\rowcolor[HTML]{FFFFFF} 
\multicolumn{2}{r|}{\cellcolor[HTML]{FFFFFF}BL1p}                   &                                                 & 25.42                                            & 5.40                                               &                                                 & 3.39                                             & 3.55                                               \\ \hline
\end{tabular}
  \caption{
The weighted total mean absolute deviations (WTMAD-2, kcal/mol) are calculated on the spin-contaminated (SC) and spin-uncontaminated (SU) reactions using various sc-/UHF-/ROHF-DFT methods. 
Number of reactions included in SC/SU are written in the parenthesis. 
$^\dag$Best performing functional in each rung among the accessed functionals in Ref.~\cite{GHBE17} are shown for comparison. 
$^\ddag$sc-\rtscan{}-D4 is calculated with Grimme's original set of parameters. (See Table S1 for the D4/DC4 parameters.)
  }
  \label{tab:15fnls_table}
\end{table}

In DC-DFT, correcting the density means eliminating (or reducing)
the density-driven error by calculating the DFA energy on the exact (or better) density instead when the sc-density is wrong.
As the exact density $n({\bf r})$ is not available in practical calculations,
very often the HF density is used in its place.
Sim et al. found that in many well-characterized cases, typical DFA errors are greatly reduced by applying HF-DFT.\cite{SSVB22,NSSB20,SVKY22,NCSB21,KSB11,KPSS15,KSB14,SKSB18,KSSB19,JS08,VPB12}
The abnormality of a given calculation depends on the property, the system, and the DFA being used.
Moreover, if a calculation is normal, removal of the density-driven error might even slightly worsen
results.   To determine when one should throw out the sc-density, the
concept of density sensitivity was introduced.
Density sensitivity \S{} can be practically quantified as\cite{KSSB19}
\begin{equation} \label{eq:S}
    \tilde{S} = |\tilde{E}[n^{LDA}]-\tilde{E}[n^{HF}]|
\end{equation}
where $\tilde{E}$ is the DFA of interest and $n^{LDA}$ and $n^{HF}$
are the electron densities obtained by LDA and HF, respectively.
A sensitivity over 2 kcal/mol provides a practical guide for when
sc-densities are problematic.  This generic rule works well
for covalent bonds in small molecules, but must be modified
for weaker bonds or bigger molecules. 
Other metrics to evaluate density sensitivity have been proposed,\cite{MH21,GT23} 
and a suitable method for the context should be chosen.

In DC(HF)-DFT, it has been suggested to replace the sc-density with the HF density in density-sensitive (DS) cases, 
which has proven very successful.\cite{SVSB22}
The commonsense assumption for why this works
is that, in cases with high density-driven errors despite small functional errors, the HF density is much closer
to the exact density than the sc-density is, at least by the energetic
measure of DC-DFT.  In the few cases where KS inversions have been sufficiently
accurate, this has been found to be true.\cite{NSSB20}  Recent work on barrier heights\cite{KSBP23,KKSG23}
argues that these are not in fact typically abnormal, based on evidence from
proxy calculations and demanding open-shell KS inversions, but leaving the
tremendous improvements in barrier heights from HF-DFT more difficult to explain.

The energy of a DS calculation
tends to vary a lot depending on different density inputs, 
and small density errors may cause large density-driven errors.\cite{KSB13} 
DC(HF)-DFT is a method where HF density is only applied to DS cases, 
and sc-density is used in density-insensitive (DI) cases.\cite{SVSB22}
On the other hand, HF-DFT is the indiscriminate use of the HF density
in {\em all} cases, regardless of density-sensitivity.   But since
DI cases greatly outnumber DS cases in large
databases, such as GMTKN55, a small (and unimportant) increase
in errors from using the HF density when inappropriate can easily
mask the large (and significant) improvement due to the HF density
in DS cases.\cite{SVSB22,SM21}

But whether one uses HF-DFT or the more nuanced DC(HF)-DFT, there is one situation that has
been left unresolved:  What do we do when the HF density is significantly spin-contaminated?
That is the subject of the rest of this paper.

\noindent
{\em Spin contamination in UHF} calculations means that the wavefunction is contaminated by higher spin states,
instead of representing a desired single spin state.\cite{S98}
The amount of spin contamination \dSt{} can be practically measured by the deviation of the spin expectation value from the exact value that should come out from a wavefunction of a pure spin state;\cite{WDHW11}
\begin{equation} \label{eq:dS2}
\textbf{\dSt{}} = \textbf{\St{}} - S_z(S_z+1) .
\end{equation}
Spin contamination can appear and be evaluated in many open-shell quantum chemistry methods,
such as in HF, DFT, second order Møller-Plesset perturbation theory (MP2), and coupled-cluster singles and doubles (CCSD), etc.\cite{S98,AJBH91,MR08}
Post-HF or double-hybrid density functional calculations using a
spin-contaminated UHF wavefunction can yield very poor results.\cite{AJBH91}
Yet, the unrestricted scheme is the most frequently used open-shell scheme in both HF and KS-DFT.
Its simple definition and ease of computation make the unrestricted scheme highly desirable.
UKS wavefunctions are less likely to be spin-contaminated
than their HF counterparts,\cite{BSA93,MTS00} which has led to less attention
to the problem of spin contamination in DFT.

It is important to distinguish our use of ROHF from those traditionally used in wavefunction calculations or in DFT.
For a wavefunction method starting from a HF calculation, spin contamination
of the starting point can lead to severe inaccuracies in any wavefunction built upon it.\cite{WDHW11} Since a perfect
method would be independent of the starting point, but imperfect methods are not, significant improvement
in the quality of the wavefunction can be achieved by removing spin contamination.   On the other hand,
there are strong arguments against removing spin contamination in DFT calculations, especially for materials.\cite{PSB95,ZM21,PCSK22}
For approximate functionals, a broken spin-symmetry solution will typically yield the best energetics,
and even the broken-symmetry densities can capture frozen fluctuations of the true ground state.
Neither of these cases applies here, as we are simply asking which HF densities yield the best energies
when approximate density functionals are applied to them (and none are self-consistent).

Many ROHF schemes or spin-projected UHF schemes
have been suggested to deal with the problem, %
which perfectly or partially remove the spin contamination through various means.
A weakness of ROHF is that it is not a uniquely defined method, nor does it provide a single set of orbitals.
This leads to difficulty in analyzing the orbital energies
or defining perturbation methods based on ROHF orbitals.\cite{J07}
There exist studies comparing orbital energies or total energies of different open-shell HF schemes,\cite{AJBH91}
but there have been no studies comparing the densities or their effect on HF-DFT energies.
Here we focus on the influence of spin-contaminated UHF density versus spin-pure ROHF density on
HF-DFT calculations and compare the results.
We call them UHF-DFT and ROHF-DFT respectively.

We need a uniquely defined ROHF scheme that differs from UHF mainly in the case of spin contamination.
There exist many combinations of ROHF coefficients\cite{PGB06}
or projected UHF schemes.\cite{AJBH91,B89,AH61,TS10} 
We have chosen the constrained-UHF (CUHF) algorithm, 
which employs parameter-free Fock matrices to mathematically constrain the spin density eigenvalues of UHF. 
This approach yields orbital energies that retain their physical significance similar to UHF, while effectively eliminating spin contamination.\cite{TS10}
The scope of CUHF can be extended as a bridge between ROHF and UHF by widening the active space of the orbitals, 
and MP2 utilizing CUHF orbitals (CUMP2) is also available.\cite{TS11}
Comparing results by varying the range of active spaces could provide a more
sophisticated study of the effect of spin contamination,
but here we have implemented the CUHF algorithm simply as a ROHF scheme.
We denote these methods as ROHF and ROMP2 in what follows.

It is important to note that the ROHF wavefunction may have lost some other features in return for the exact spin eigenvalue,
and other errors may be inherent in its density.
However, Fig.~\ref{fig:firstpic} and further discussion below shows that, at least when spin contamination in UHF is severe,
simply replacing by ROHF can effectively reduce the errors in UHF-DFT induced by spin contamination.
We will define every open-shell system as spin-uncontaminated (SU) or spin-contaminated (SC),
and compare the performance of UHF- and ROHF-DFT in each.
Typically, a UHF wavefunction with \dSt{} over 0.1 \cite{WDHW11} or \dStp{} over 10~\% \cite{RHDB20} is
considered severely spin-contaminated.
As seen in Fig. S1, the conventional criterion 0.1 works well for this study,
so we divide SC/SU by the same criterion 0.1.

Before we continue to the results, we must sound a note of caution.
The spin density plots of UHF and ROHF in Fig.~\ref{fig:firstpic} are provided simply to convey the concept.
In fact, in most interesting cases, we find it impossible to decide which is a `better' density by simple
inspection of such plots.
`Correcting' the density in DC-DFT means to reduce the density-driven error,
but this does not directly translate into  a visually favorable density.
Very tiny features in densities can yield significant differences in energies.
Two densities may appear remarkably similar, but have substantial different XC
energies with a given functional.  On the other hand, densities that differ
significantly in some region might have almost identical energies.
Moreover, which is which depends on the functional being applied.
Thus, within DC-DFT, contour plots of densities and density differences, while useful,
can never substitute for accurate calculation of density-driven errors.



Here, we study simply the effect of using either HF
density on open-shell cases. Of the 1505 numbers in the GMTKN55
database, about 30~\% (in fact, 430) contain an open-shell species.
We call these G55o, for brevity.  Of these cases, about 10~\% (in fact, 46)
are spin-contaminated (i.e., above the 0.1 level).  Table 1 gives results
for 13 different functional approximations, comparing errors when
self-consistent, UHF, and ROHF densities are used.   
Four functionals indicated by a dagger are the best performing functional in each rung (GGA/mGGA/hybrid/double-hybrid) among the ones assessed in Ref.~\cite{GHBE17}, 
and their self-consistent results are given for comparison.

We compare the performance by the weighted total mean absolute deviation (WTMAD-2), proposed together with the GMTKN55 database,\cite{GHBE17} instead of the conventionally used mean absolute errors (MAE). 
WTMAD-2 compares errors in different subsets by giving weights depending on their reference energies. 
The average relative absolute reference energies (MAR) of each of the 55 test sets in GMTKN55 
vary from 0.58 kcal/mol (RG18) to 654.26 kcal/mol (DIPCS10).
To fairly compare the relative energies of different test sets, 
the authors proposed two weighted total mean absolute deviation (WTMAD) schemes for statistical analysis. 
Because the first scheme WTMAD-1 weights each test set arbitrarily, 
here we use the second scheme WTMAD-2. 
In WTMAD-2, the weight is set by the ratio between 56.84 kcal$\cdot$mol$^{-1}$ and the mean absolute reference energy (MAR) for the respective test set. 
\begin{equation} \label{eq:WTMAD-2}
    \WTMAD = \frac{1}{\sum_i^{55}{N_i}} \cdot \sum_i^{55}{N_i} \cdot \frac{56.84 kcal\cdot mol^{-1}}{\mar} \cdot \mae_i 
\end{equation}
Using this weighted scheme, small relative energies such as weak non-covalent interactions have more influence on the performance.

For spin-contaminated reactions, use of the ROHF density yields better
energetics than the UHF density, for {\em every} functional
listed. 
The errors are reduced by at least 30~\% and sometimes up to 70~\%. 
Comparing self-consistent versus ROHF densities on spin-contaminated cases, 
ROHF densities reduce the errors for most of the functionals, except for BLYP and M06-2X. 
For BLYP, the difference is about 1 kcal/mol of WTMAD-2, 
and for M06-2X, it is about 4.5 kcal/mol. 
The behavior of HF densities on Minnesota functionals and hybrids is not expected to be consistent 
because they are empirically fitted to reduce the total error, without separating the density-driven errors from functional errors.
Furthermore, M06-2X includes 54\% of HF exchange, which could double-count the effect of exact exchange. 
The ROHF-DFT methods even outperforms the four functionals, chosen in Ref.~\cite{GHBE17} as the best performing functional on the GMTKN55 database in each rung.

The last two lines of Table~\ref{tab:15fnls_table} are designed to test ROHF for two DC functionals,
i.e., functionals designed to be used on HF densities.
These are 
HF-\rtscan{}-DC4 and BL1p.
In the former case, we also compare to sc results with \rtscan{}-D4, using the standard
D4 parameters for \rtscan{}.   We describe these below.

The recently published HF-\rtscan{}-DC4 was proposed as
a general and practical protocol, taking advantage of insights from DC-DFT.\cite{SVKY22}
The method has shown remarkable performance in calculating the properties
of various systems containing water and describing non-covalent interactions of biomolecules.
When this method was introduced, three important elements of DC-DFT were highlighted:
(a) The HF density reduces density-driven errors,\cite{DLPP21} (b)
 \rtscan{} is used instead of SCAN,
thereby avoiding grid convergence issues, and (c) a DC-DFT-based parameterization
of Grimme's D4 dispersion correction handles the dispersion energies. 
Here, we relabel the method as UHF-\rtscan{}-DC4, since the UHF density was used.
For comparison, we now introduce ROHF-\rtscan{}-DC4, which uses ROHF densities instead
for open-shell cases. 
To compare the influence of the two different HF densities,
we used the same DC4 parameters in both methods without refitting the parameters. 
The number of open-shell density-insensitive cases included in the original training set is small, in order to avoid spin contamination in UHF calculations.\cite{SVKY22} 
Refitting the parameters considering the cases would yield a slightly better performance.
In the following, (sc-)\rtscan{}-D4 and UHF-/ROHF-\rtscan{}-DC4 results are investigated.
(The parameters and references for D4 and DC4 are listed in Table S1.)

The other DC-DFT method implemented is BL1p, a HF-DFT-based one-parameter 
double-hybrid functional.
BL1p was introduced as a prototype to demonstrate the importance of DC-DFT
when fitting empirical parameters.   Since most fitting uses the total energy error,
functional and density-driven errors are entangled, reducing the accuracy of the
fit.  BL1p
demonstrated the higher accuracy that could be achieved simply by removing
DS cases from the training set.\cite{SVSB21}
The BL1p energy is calculated on the HF density and utilizes
the MP2 energy obtained from the HF orbitals;
\begin{multline}
E\xc^{\sss BL1p}[n^{\sss HF}]
= E\xc^{\sss BLYP} + \alpha(E\x^{\sss HF} - E\x^{\sss B88}) \cr
+ \alpha^2 (E\c^{\sss MP2} - E\c^{\sss LYP}).
\label{eq:bl1p}
\end{multline}
\noindent
All open-shell calculations used UHF in the
original paper, as seen in the parameter scan result in Fig. S2.
To compare the two open-shell HF methods, 
we call the original form using UHF density and UMP2 energy UBL1p,  
and a new form using ROHF density and ROMP2 energy ROBL1p. 
The single empirical $\alpha$ parameter in UBL1p was set by 0.82 by scanning through the AE6 dataset, 
a small representative subset in the Minnesota database\cite{LT03} containing atomization energies of 6 molecules.
The same process was run with ROBL1p, but the difference in the optimal value was negligible. 
(See Fig. S2.)
Therefore, the same value, 0.82, was used for ROBL1p throughout.

\begin{figure}[!htbp]
\centering
\includegraphics[width=\columnwidth]{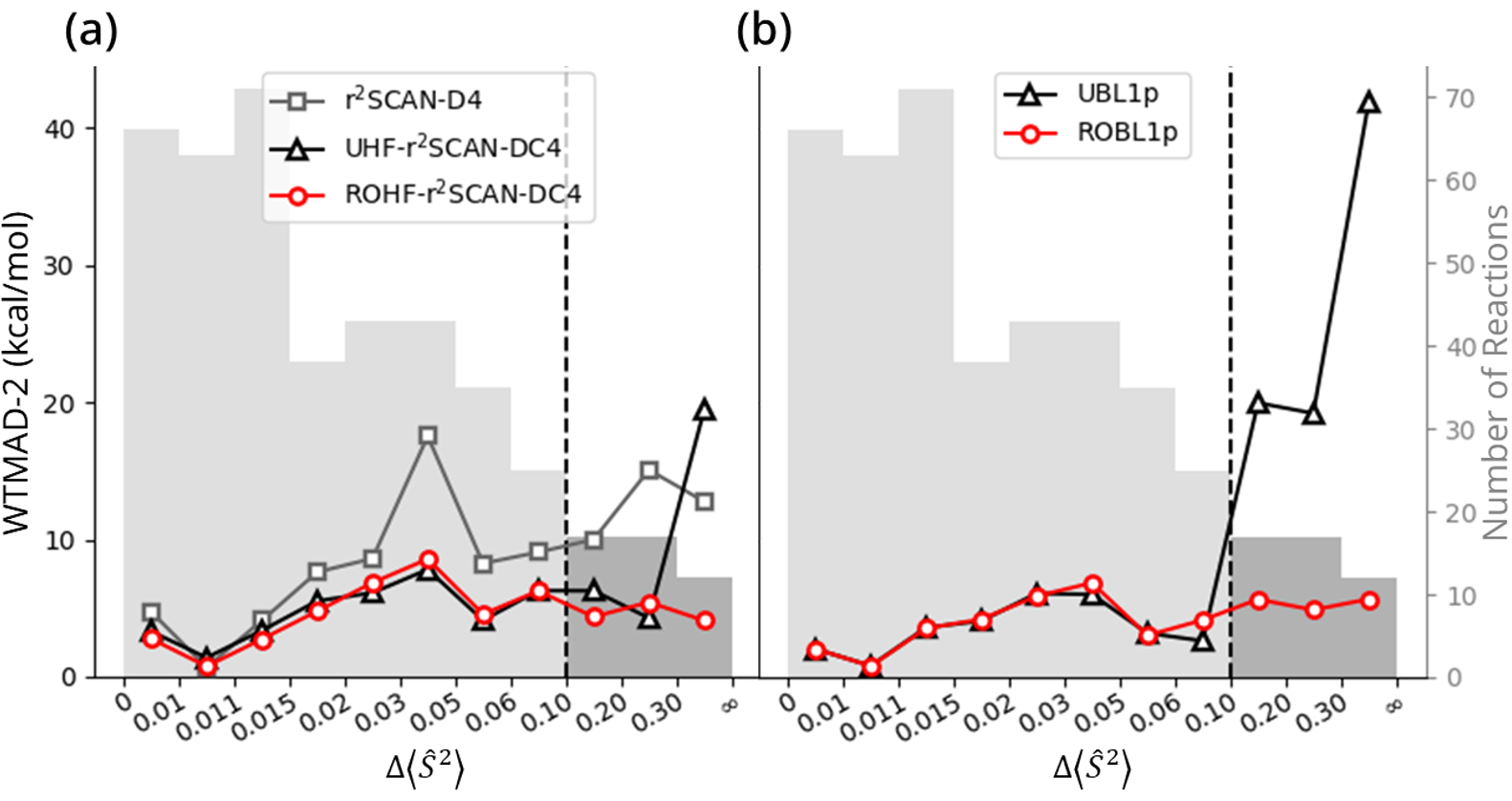}
\caption{
The weighted total mean absolute deviations (WTMAD-2, kcal/mol) of different methods for varying values of \dSt{} in G55o are presented. 
In (a), sc-\rtscan{}-D4, UHF- and ROHF-\rtscan{}-DC4 are plotted in gray, black and red, respectively. 
In (b), UBL1p and ROBL1p are plotted in black and red, respectively. 
Gray bars represent the number of reactions included in the range of \dSt{}. 
The vertical dashed line at \dSt{}=0.1 divides the data into two categories: spin-uncontaminated (SU, left) and spin-contaminated (SC, right).
Note that the ranges were arbitrarily chosen to achieve a similar number of reactions within each range.
}
\label{fig:MAEs_per_dS2}
\end{figure}

The DC-DFT rows of Table~\ref{tab:15fnls_table} shows the mean performance of the five methods; sc-\rtscan{}-D4, U/ROHF-\rtscan{}-DC4, and U/ROBL1p.
In both cases, ROHF yields much improved results for the spin-contaminated set.
ROHF slightly worsens the SU cases for BL1p, but the effect is so small that its overall performance
is still improved relative to UHF.   Interestingly, ROHF-\rtscan{} performs
slightly better than ROHF-\rtscan{}-DC4, for spin-contaminated cases.  
This could be an error cancellation due to the lack of dispersion, 
or because the spin-contaminated cases were not heavily considered when fitting the DC4 parameters. 
Either explanation would be interesting to study further.
Still, ROHF-\rtscan{}-DC4 works better than sc-\rtscan{}-D4, which was not fitted based on DC-DFT.

How sure are we that the improvement for spin-contaminated cases is not accidental?
For a more in-depth analysis, the WTMAD-2 of the original UHF-based and ROHF-based methods are
compared in Fig.~\ref{fig:MAEs_per_dS2}, as a function of the level of spin contamination. 
The two HF-\rtscan{}-DC4 schemes exhibit similar performance for  \dSt{}<0.1, i.e., for spin-uncontaminated cases.
The MAE using the UHF density jumps tremendously when \dSt{}>0.30, becoming much larger than that of ROHF.
UBL1p and ROBL1p exhibit similar trends to HF-\rtscan{}-DC4, but display an even larger difference in the SC region. 
This discrepancy can be attributed to the inclusion of the MP2 part in BL1p, which is even
more susceptible to contamination in the wavefunction.

\begin{table}[!htbp]
\centering
\includegraphics[width=0.9\columnwidth]{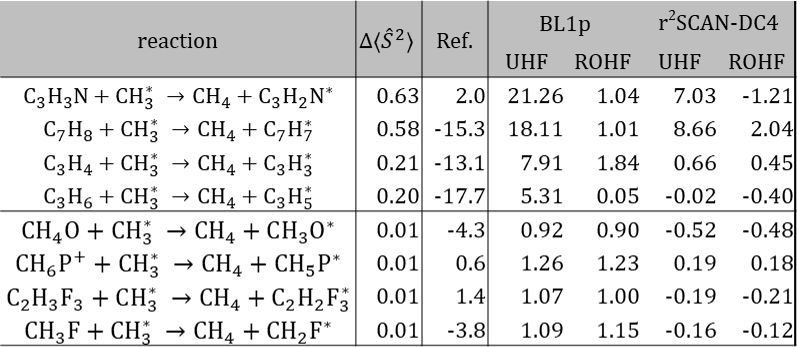}
\caption{
Errors for the four most extreme reactions from the RSE43 data set: the highest (SC, top) and the lowest (SU, bottom) spin contamination,given in the \dSt{} column.
The Ref. column shows the reference reaction energies from Ref.~\cite{GHBE17}. 
The four columns on the right show the errors of each method calculated by reaction energies minus reference energies in kcal/mol.
Results using BL1p are also shown in the graphical abstract.
}
\label{tab:RSE43_reactions}
\end{table}

Now, we focus on some specific dramatic examples.
Table~\ref{tab:RSE43_reactions} shows the four most and the four least spin-contaminated cases in the RSE43 dataset,
a subset of the GMTKN55 database consisting of radical stabilization energies. 
In the four highly spin-contaminated cases, UBL1p yields high errors by calculating the MP2 energy based on a highly spin-contaminated UHF wavefunction.
Especially for the two reactions where \dSt{} is over 0.5, the UHF error is $\sim$20 kcal/mol, but only $1$ kcal/mol in ROHF.
On the other hand, the SU cases show very small differences between UHF and ROHF. 
A similar trend occurs for HF-\rtscan{}-DC4, but the failure in spin-contaminated reactions is larger in UBL1p than in UHF-\rtscan{}-DC4. 
Menon and Radom showed that double-hybrid functionals are less likely to be affected by spin contamination, compared to pure UHF and UMP.\cite{MR08}
But BL1p is evaluated on the HF density and includes the UMP2 energy, so the errors induced by spin contamination are larger. 

Therefore, one should be cautious when using the HF wavefunction or density,
and should avoid using a highly spin-contaminated UHF wavefunction. 
Severe spin contamination in UHF could also indicate a multi-reference character of the system, 
and in that case, applying ROHF might not help.
Moreover, here we propose and test the well-known ROHF scheme as a cheap and simple solution, 
but there could be other alternatives that treat spin better than either UHF or ROHF.


{\em DC(HF)-DFT avoiding spin contamination:}
Now we combine the above discussion with the DC(HF)-DFT protocol. 
Previously the protocol was to check the density sensitivity
and decide whether to use the HF density or not. 
In cases of UHF strong spin contamination, one simply reverted to
the self-consistent DFT density instead.

\begin{figure}[!htbp]
\centering
\includegraphics[width=\columnwidth]{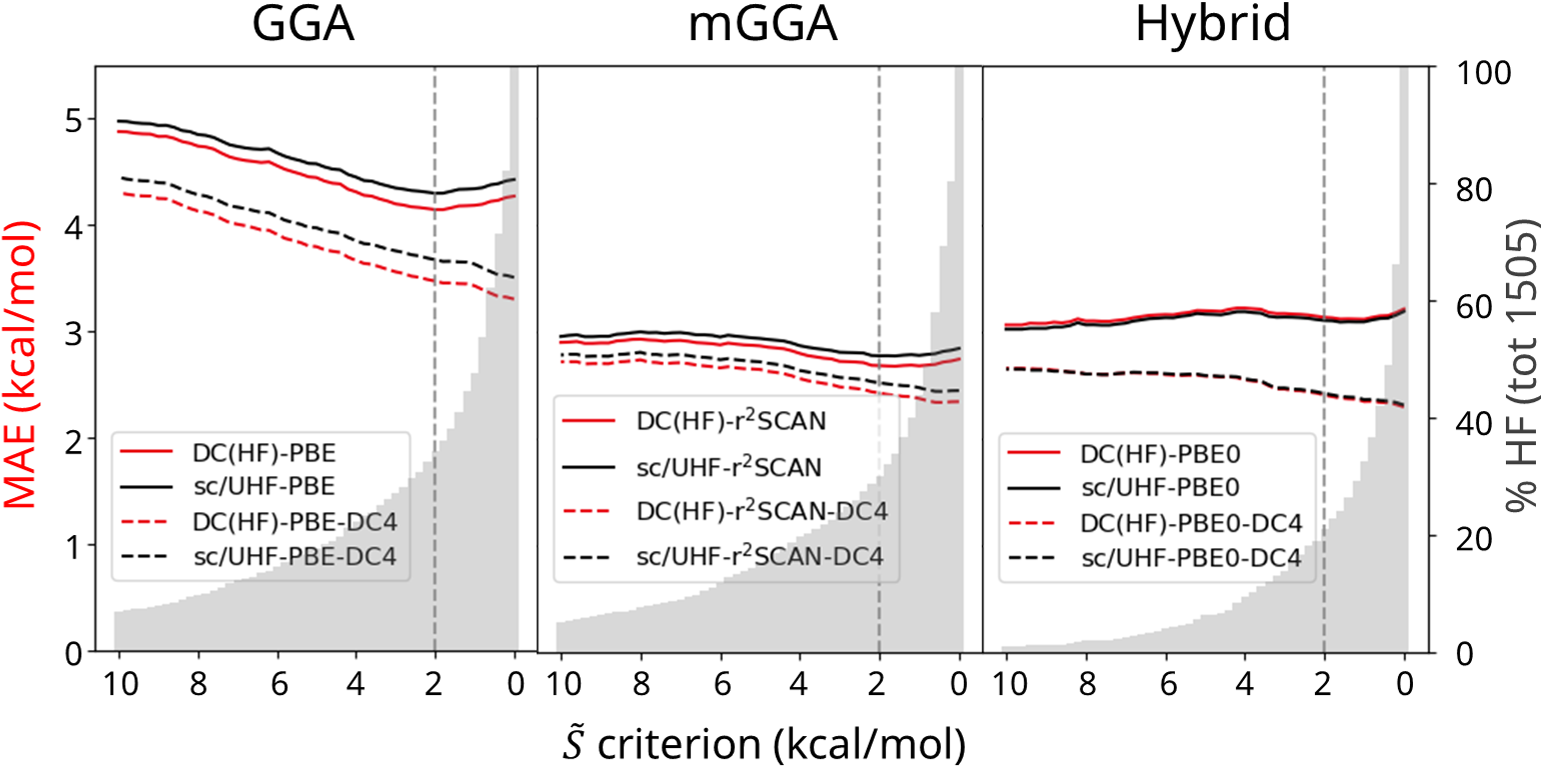}
\caption{
\S{} criterion was scanned from 0 to 10 on all reactions in the GMTKN55 database. For a certain criterion, the density for each reaction is chosen between HF and self-consistent, depending on whether its \S{} value is over the criterion or not. In the sc/UHF-DFT (black lines) methods, UHF densities are used without consideration of spin contamination. In the DC(HF)-DFT (red lines) methods, ROHF densities are used when UHF is spin-contaminated. Results are shown for GGA(PBE), mGGA(\rtscan{}), and Hybrid(PBE0). The vertical dashed line indicates \S{} criterion at 2 kcal/mol. Gray bars show the percentage of HF densities chosen at each criterion.}
\label{fig:S_scan}
\end{figure}

Our ROHF results dictate an alternative.
Before calculating density sensitivity, check the UHF spin contamination.
If the value is over the 0.1 criterion, UHF should be replaced by ROHF. 
We define \S{} as \S{}$_{\text{U}}$ or \S{}$_{\text{RO}}$ from eq.~\ref{eq:S} using UHF and ROHF densities, respectively.
\begin{equation}
\tilde{S} = \begin{cases}
    \tilde{S}_{\text{RO}}, & \text{if SC ($\Delta \langle \hat{S}^2 \rangle \geq 0.1$)}, \\
    \tilde{S}_{\text{U}}, & \text{otherwise ($\Delta \langle \hat{S}^2 \rangle < 0.1$)}.
\end{cases}
\end{equation}
As mentioned in the sensitivity criterion scan results in Fig.~\ref{fig:S_scan}, 
we set the criterion of density-sensitivity as 2 kcal/mol, the previously suggested practical criterion.\cite{SVSB22}

Finally, after the appropriate density has been chosen, 
a dispersion correction should be added for functional error correction.
Dispersion corrections are vital to correctly describe 
non-covalent interactions or long-range interactions, and
parameters should be optimized based on DC-DFT principles, so
as not to spoil the dispersion correction by the density-driven error.\cite{KSSB19} 
For example, DC4 can be used, which is a variation of Grimme's D4 dispersion correction\cite{CEHN19} parameterized by Song et al.\cite{SVKY22} to create HF-\rtscan{}-DC4.

\begin{figure}[!htbp]
    \centering
    \includegraphics[width=\columnwidth]{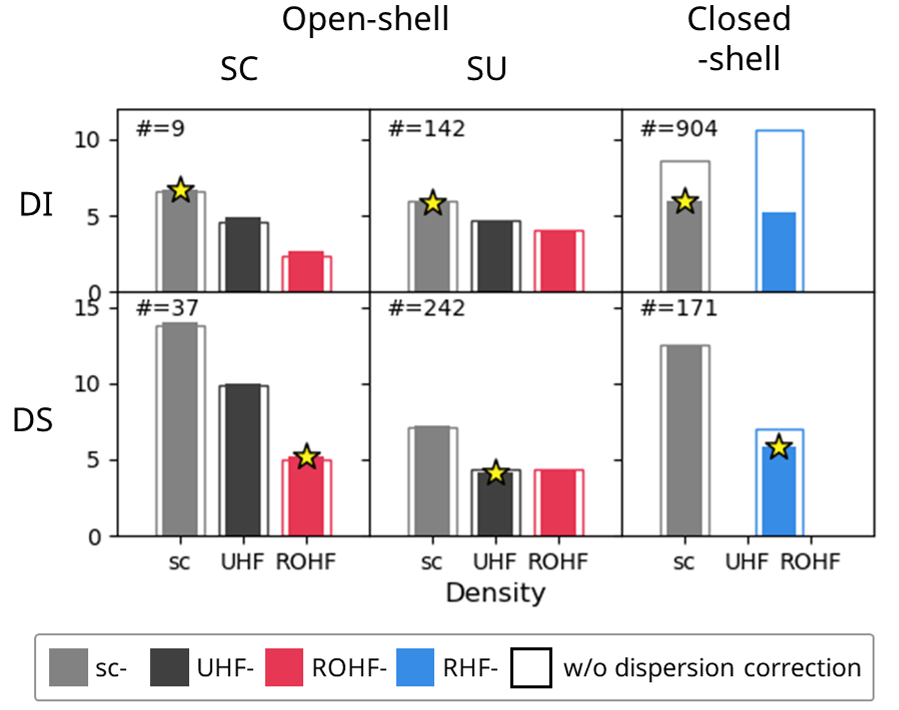}
    \caption{
    The weighted total mean absolute deviations (WTMAD-2, kcal/mol) of \rtscan{} calculated on different densities are plotted, for the reactions in GMTKN55 grouped by open(SC/SU)/closed and DI/DS. The number of reactions contained in the groups are written in the upper right of each graph. Gray/blue/black/red bars indicate WTMAD-2 values of sc-/RHF-/UHF-/ROHF-\rtscan{}, and filled/empty bars are with/without dispersion correction, which is D4 for sc-DFT and DC4 for HF-DFT methods. The yellow star points the DC(HF)-DFT-DC4 MAEs, chosen for each groups following the suggested recipe. The same figure for other functionals are provided in Fig. S3.
    }
    \label{fig:MAE_venns}
\end{figure}

Figure~\ref{fig:MAE_venns} shows the WTMAD-2 values of \rtscan{} calculated on 
self-consistent and HF densities,
categorized by closed/open-shell (SU/SC) and DI/DS.
(Other functionals are shown in Fig. S3.) 
Yellow stars indicate the densities chosen by the recommended DC(HF)-DFT scheme,
and in most cases, the yellow stars follow the lowest energies.
For SU-DI cases, HF and sc-densities show similar performances as expected. 
The improvements of HF density over sc-density for SU-DS cases also match the previous studies of DC-DFT. 
ROHF densities clearly reduce the error for the spin-contaminated cases, 
which shows that the UHF-DFT error in the region is due to the spin-contamination of the UHF wavefunction. 

Figure ~\ref{fig:S_scan} shows the \S{} criterion scan results using three different functionals: PBE, \rtscan{}, and PBE0. Red dashed lines show the DC(HF)-DFT scheme suggested above. 
For DC(HF)-DFT without dispersion correction (red solid lines), local minima appear near the conventional DS criterion of 2~kcal/mol.
This means that using HF densities only
for density-sensitive cases (and self-consistent densities for all others)
gives the best results. However, addition of DC4 eliminates this minimum
and the to switch densities, i.e., errors are least when HF densities are used consistently. 

In general, it has been recommended to use the HF density only when
reactions are density-sensitive, 
but when dispersion corrections (fit correctly following the scope of DC-DFT) are
included,
using the HF densities always yields the best performance. 
Among these particular examples shown in Fig.~\ref{fig:MAE_venns} and \ref{fig:S_scan}, \rtscan{}-DC4 with ROHF density showed to be the best in all six categories, regardless of \S{} (i.e., criterion zero in Fig.~\ref{fig:S_scan}). 
Therefore, one could always use ROHF density for this case, but with a caveat. 
For spin-contaminated and density-insensitive cases, the WTMAD-2 error reduction is noticeable when using ROHF densities, but this is due to the inclusion of many reactions with relatively small reference energies and, therefore, large weights in the WTMAD-2 scheme, such as the radical stabilization energy subset (RSE43). The MAE difference between sc- and UHF/ROHF densities is smaller than 1.4 kcal/mol.
Further study may yet yield better densities, but the ROHF density is a practical remedy for spin contamination at present.


\textbf{In summary}, our study highlights the importance of
considering spin contamination in open-shell HF-DFT calculations. 
HF-DFT has received a lot of attention recently due to its cost-effective nature and significant energetic improvements in DFT calculations. 
The method involves calculating DFT energies on HF densities instead of their self-consistent ones. 
Based on this success, HF-\rtscan{}-DC4 has been developed and has shown remarkable performance in challenging systems like water. 
However, previous performance studies of HF-DFT have avoided
the issue of spin contamination.
For example, it was briefly discussed in the context
of DC(HF)-DFT that HF-DFT should only be applied where
density sensitivity is high and spin contamination is low.\cite{SVSB22}

In this study, we have provided performance studies of
two different open-shell HF densities using various density functional approximations.
For spin-contaminated cases, ROHF densities reduced WTMAD-2 errors relative
to self-consistent densities for all
three types of functionals, including GGA, mGGA, and hybrid, while giving an
only slightly higher WTMAD-2 in uncontaminated cases.
The double-hybrid HF-DFT functional BL1p suffered most severely from spin contamination
in UHF.
While the two HF densities showed similar performance in open-shell SU cases, 
treating the spin-contamination by using ROHF densities showed clear improvements in spin-contaminated cases
with WTMAD-2 values varying from 25 kcal/mol for UBL1p to 5 kcal/mol for ROBL1p.
Even for the less pronounced HF-\rtscan{}-DC4, 
ROHF densities reduced the spin-contaminated WTMAD-2 by about a factor of 2
relative to 
UHF densities. 
These results highlight the importance of considering spin
contamination in open-shell HF-DFT calculations.

Furthermore, we combine the results with the DC(HF)-DFT concept
to underscore the importance of caution when applying DC-DFT
in systems with spin contamination, 
and provide guidance for handling open-shell systems.
We hope that these findings extend the applicability of HF-DFT
to a wider range of systems,  
and provide valuable insights into open-shell computations.

\section*{Computational Details}

The GMTKN55 database includes 5 subsets, and the details are presented in Table S2.
All reference energies, geometries of systems, and sc-DFT results except for \rtscan{} are from Ref.~\cite{GMTKNweb}.

Reactions that contain one or more open-shell systems are marked as open-shell reactions, otherwise closed-shell reactions. 
For open-shell reactions, all constituent open-shell components are calculated with the same HF method. That is, we do not mix UHF and ROHF densities in one reaction.
We define \dSt{} of a case (reaction or energy difference) by the highest \dSt{} value among all constituent components, and \S{} by eq. 4 where $\tilde{E}$ is the corresponding energy difference. 
In closed-shell systems, all HF calculations are carried out using the restricted form (RHF), which has no bearing on the open-shell HF comparison. 

All cases can be classified according to whether they are density sensitive/insensitive, and spin-contaminated/uncontaminated.
Conventionally, a UHF wavefunction with \dSt{} over 0.1\cite{WDHW11} or its percentage over 10~\%\cite{RHDB20} has been considered spin-contaminated. 
In this work, we follow the former definition, 
labeling a reaction as spin-contaminated (SC) if its \dSt{} is over 0.1, and spin-uncontaminated (SU) otherwise. 
For density sensitivity, we follow the criterion 2 kcal/mol from Sim et al.,
and label a reaction as density sensitive (DS) if its \S{} is over 2 kcal/mol and density insensitive (DI) otherwise.\cite{SSB18}
The sensitivity value depends on the functionals, and therefore the number of reactions included in DS/DI groups differ.

All HF and DFT calculations are performed via the Python-based Simulations of Chemistry Framework,\cite{pyscf} utilizing customized Python codes for CUHF. 
The Ahlrichs def2-QZVPPD basis set\cite{def2-1,def2-2} was used for all calculations. 
The methods analyzed are the sc-/UHF-/ROHF-DFT with 4 generalized gradient approximations (GGAs) (BLYP\cite{B88,LYP88,BLYP}, RPBE\cite{RPBE}, PW91\cite{PW91}, PBE\cite{PBE}), 4 meta-GGAs (TPSS\cite{TPSS}, M06L\cite{M06L}, SCAN\cite{SCAN}, \rtscan{}\cite{FKNP20}), 5 hybrids (B3LYP\cite{B93,B3LYP}, TPSSh\cite{TPSSh}, PBE0\cite{PBE0,pbe0-2}, M06\cite{M06}, M06-2X\cite{M06}), and two fully HF-DFT methods HF-\rtscan{}-DC4\cite{SVKY22} and BL1p\cite{SVSB21}.

\section*{Supporting Materials}

%
%
%
%

The Supporting Information are available.

D4/DC4 parameters, numbers of open-shell/spin-contaminated reactions contained in each categories of GMTKN55, \dSt{} criterion scan results, BL1p $\alpha$ parameter scan results, and Fig.~\ref{fig:MAE_venns} for other functionals. (PDF)

\noindent Raw data of all calculations performed in this work. (csv)

\section*{Acknowledgment}
We are grateful for support from the National Research Foundation of Korea (NRF-2020R1A2C2007468) and Korea Environment Industry \& Technology Institute (KEITI) through “Advanced Technology Development Project for Predicting and Preventing Chemical Accidents” Program funded by Korea Ministry of Environment (MOE) (RS-2023-00219144).
K.B. acknowledges support from NSF grant No.CHE-2154371.

\clearpage






\end{document}